\documentclass[aps,
reprint,
amsmath, 
amssymb,
nofootinbib,
raggedbottom,
superscriptaddress]{revtex4-2}
\usepackage{tikz}
\usetikzlibrary{decorations.pathmorphing, decorations.markings}
\usepackage{graphicx} 
\usepackage{preamble}
\usepackage{color}
\usepackage[dvipsnames]{xcolor}
\usepackage{cancel}
\usepackage{orcidlink}
\usepackage{soul}

\usepackage{mathtools}

\usepackage{bm}

\begin{document}

\title{Quantization of Beta Functions in Self-Dual Backgrounds 
\\ and  Emergent Non-Commutative EFT}

\author{Mithat \"{U}nsal \orcidlink{0000-0002-4875-9978}}
\affiliation{Department of Physics, North Carolina State University, Raleigh, NC 27607, USA \looseness=-1}

\begin{abstract}
We investigate the renormalization group flow and beta functions of   Yang-Mills theory and adjoint QCD  in a strong, stable, self-dual background field $F$. In deep UV, theory runs according to the standard beta function, $\beta_0$.
Treating the background as a superselection sector, we find that the theory abelianizes below the scale $\sqrt{F}$ and remains strictly abelian in the deep infrared. In the intermediate weakly-coupled regime ($\Lambda_{\rm YM} \ll \mu \lesssim \sqrt{F}$), the gauge coupling remarkably continues to run despite the absence of propagating charged degrees of freedom. Because all non-zero Landau levels decouple, this running is driven exclusively by exact zero modes, resulting in an integer-quantized beta function coefficient, $\widetilde \beta_0$. Finally, we conjecture that this abelian dynamics is governed by an emergent non-commutative effective field theory that is  free of pathological UV/IR mixing.  
\end{abstract}

\maketitle

\subsection{Superselection sectors in QCD}
Developing reliable analytical techniques to study the dynamics of gauge theories directly on $\mathbb{R}^4$ remains a central challenge. While exact kinematic constraints from generalized global symmetries and 't Hooft anomalies  provides insights into the possible infrared vacuum structures \cite{tHooft:1979rat, Gaiotto:2017yup}, dynamical questions often require different tools \cite{Seiberg:1994rs, Polyakov:1975rs, Unsal:2007jx}. 

In this work, we show that Yang-Mills theory on a large, constant self-dual background provides such a toolbox, complementary to earlier studies using compactification 
\cite{Unsal:2007vu, Unsal:2007jx, Tanizaki:2022ngt, Poppitz:2021cxe}.
Although this is a well-known set-up \cite{Leutwyler:1980ma}, we introduce a conceptual shift to the standard paradigm, viewing it as a superselection sector.  This will lead to some striking results in gauge dynamics on Euclidean $\mathbb{R}^4$. 

The study of Yang-Mills theory in a constant field strength background $F_{\mu\nu}$ is almost as old as asymptotic freedom \cite{Savvidy:1977as, Nielsen:1978rm, Nielsen:1980sx, Leutwyler:1980ma}. In these historical works, the background field is typically identified with the renormalization group scale $\mu \sim \sqrt{F}$, and the running coupling and beta function are derived at this background field scale. However, this procedure overshadows the dynamical behavior of the theory at a generic scale $\mu$ independent of the background $\sqrt{F}$. 

It is well-known that Yang-Mills theory in a purely magnetic constant background, e.g., $F_{12}=F\frac{\sigma_{3}}{2}$, suffers from the Nielsen-Olesen instability. The fluctuation operator develops a negative eigenmode (which can be interpreted as a negative "mass squared" $M_{W}^{2}=-F$ for the charged gluon $W^{\pm}$), which manifests itself as an imaginary contribution in the 1-loop potential, leading to the instability of the background \cite{Nielsen:1978rm, Nielsen:1980sx}.

In contrast, the first-order self-duality equations in Yang-Mills theory,
$$F_{\mu\nu} = *F_{\mu\nu}\,,$$
admit perfectly stable backgrounds. These equations possess \textit{i)} local instanton solutions, and \textit{ii)} global constant self-dual solutions. While the former have finite action, the latter possess only a finite action density. As demonstrated by Leutwyler \cite{Leutwyler:1980ma}, unlike the purely magnetic background (which suffers from Nielsen-Olesen tachyonic instabilities), the constant self-dual background is classically and quantum mechanically stable; the fluctuation operator possesses exact zero modes and positive modes only.

Thanks to this stability, it makes perfect sense to study the dynamics of the theory at a fixed self-dual background $F_{\mu\nu}$, but at a generic sliding scale $\mu$. In our setup, the background is not viewed merely as an RG scale, as was done historically \cite{Savvidy:1977as, Nielsen:1978rm, Nielsen:1980sx, Leutwyler:1980ma}. Rather, it is treated as a distinct superselection sector. This can equivalently be interpreted as imposing a non-trivial boundary condition on the gauge fields at infinity, $F_{\mu \nu}(|x| \rightarrow \infty) = F_{\mu \nu}$, contrasting with the standard boundary conditions where $F_{\mu \nu}(x)$ must decay faster than $1/|x|^2$. Because a constant self-dual background carries infinite total action, it does not contribute to the standard path integration over the trivial vacuum, but it rigorously defines an isolated superselection sector of the full theory.

\begin{figure}[h]
\vspace{-.3cm}
    \centering
    \includegraphics[width=0.45\textwidth]{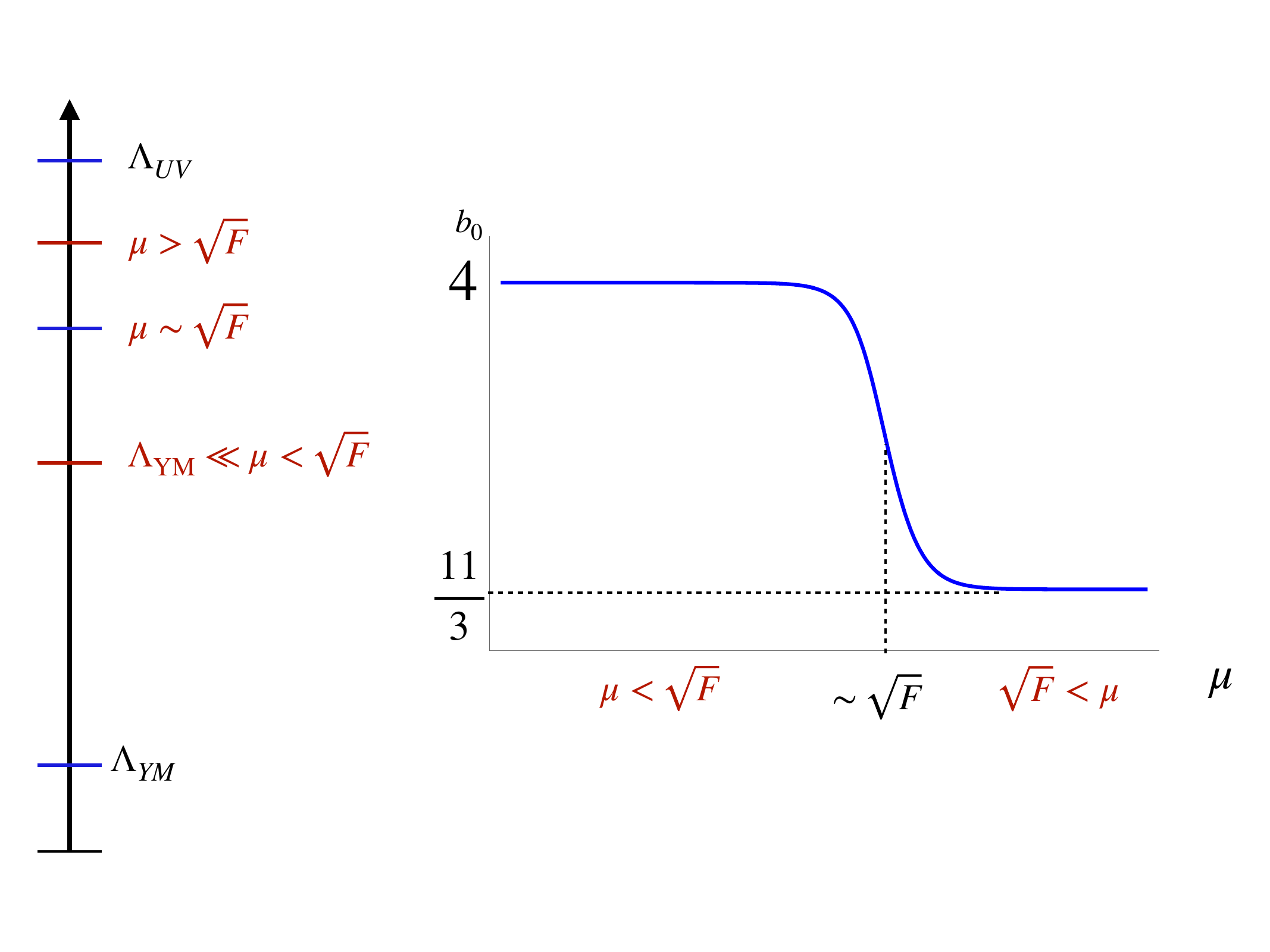}
    \vspace{-1.0cm}
    \caption{Scales in the constant self-dual background and structure of the beta function coefficient. 
 }   \vspace{-0.3cm}
    \label{fig:scales}
\end{figure}

We can therefore examine the dynamics of gauge theories within these fixed superselection sectors at a generic scale $\mu$. Our approach is conceptually analogous to studying the renormalization of Seiberg-Witten theory \cite{Seiberg:1994rs} at a fixed value of the moduli field (labeled by the adjoint Higgs scale $\langle \Phi \rangle = a\sigma_{3}/2$) across generic scales $\mu$, along the lines of Gorsky et al.\ \cite{Gorsky:1995zq, Lerche:1996xu}. The physical outcomes in our setup, however, are strikingly different.

In this work, our goal is to initiate a study of the perturbative and non-perturbative dynamics of the theory within this superselection sector, considering $F \gg \Lambda_{\rm YM}^2$. We have three main novel results:
\begin{itemize}
    \item At scales $\mu$ lower than $\sqrt{F}$ but still much larger than $\Lambda_{\rm YM}$, $SU(2)$ gauge dynamics abelianizes to $U(1)$. ($SU(N)$ generically abelianizes to $U(1)^{N-1}$.) 
    Charged gluons and matter acquire a discrete spectrum dictated by Landau levels, both in the 1-2 and 3-4 planes. Neutral components (e.g., Cartan photons) retain a continuous spectrum. 
    
    \item In the deep UV, $\mu \geq \sqrt{F}$, the beta function coefficients are the standard result, e.g., 
    \begin{equation}
        \beta_0 = \frac{11}{3}{N} \qquad (\mu \geq \sqrt{F})
    \end{equation}
    But at scales $ \Lambda_{\rm YM} \ll \mu \lesssim \sqrt{F} $, despite the fact that the theory abelianizes, and despite that there are no propagating charged modes remaining, the beta functions continue to run. In Yang-Mills (and generally in QCD),  beta function coefficients become \textit{integer quantized}:  
      \begin{equation}
        \widetilde{\beta}_0 = 4N \qquad  (\Lambda_{\rm YM} \ll \mu \lesssim \sqrt{F})
    \end{equation}
    
    \item We conjecture that at energies of order $\mu$, where $ \Lambda_{\rm YM} \ll \mu \lesssim \sqrt{F} $, the $SU(2)$ Yang-Mills theory is described by an {\it effective} non-commutative $U(1)$ theory. Unlike standard non-commutative $U(1)$ Maxwell theory, which is pathological due to UV/IR mixing \cite{Minwalla:1999px, Ruiz:2000hu}, our construction provides a healthy theory. In this sense, the construction may also be viewed as a physical realization of \cite{Grosse:2004yu, Langmann:2002cc} in gauge theory. 
\end{itemize}

\subsection{Structure of Beta Function and quantization} 

We now discuss the renormalization group in $SU(2)$ gauge theory in a constant self-dual background, where 
\begin{align}
    F_{12}= F_{34}=F \frac{\sigma_3}{2}
    \label{bgrd}
\end{align}
As shown in Fig.~\ref{fig:scales}, we can consider three weakly coupled regimes:
\begin{itemize}
    \item [i)] Deep UV ($\sqrt{F} < \mu < \Lambda_{\rm UV}$)  
    \item [ii)] Background field scale ($\mu \sim \sqrt{F}$)
    \item [iii)] Abelianized regime ($ \Lambda_{\rm YM} \ll  \mu <  \sqrt{F}$)
\end{itemize}

At energies lower than $\sqrt{F}$, only the Cartan components commute with the background. In particular, all charged gluons  $W^{\pm}$ and matter fields acquire a Landau level structure 
that depends on the  background \eqref{bgrd},  universal orbital contribution, and their spin. 

The classical action is given by 
\begin{align}
   S_{\rm YM} = \frac{1}{2g_0^2} \int d^4x \; {\rm tr} ( F_{\mu\nu}^2) + {\rm matter \; fields} 
\end{align}
where the matter sector is composed of $n_f^m$ adjoint Majorana fermions, and $n_{sc}^{\rm r}$ adjoint real scalars. The Euclidean one-loop effective action $\Gamma_{\rm eff}[A]$ is obtained by expanding the classical action around a classical background $A_\mu$ and integrating over the   quantum fluctuations. 
The fields transform under $SU(2)_L \times SU(2)_R$ as: 
\begin{align}
    {\rm background } \;  F:&  \qquad (1, 0)  \cr
    a_{\mu}:&  \qquad  \Big(\frac{1}{2}, \frac{1}{2}\Big) \equiv S_{+} \otimes S_{-} \cr 
    {\rm Fermions}: &  \qquad  \Big(\frac{1}{2},0\Big) \oplus \Big(0, \frac{1}{2}\Big)  \equiv  S_{+} \oplus S_{-}  \cr 
     {\rm Scalars}: &  \qquad  (0,0)
\end{align}
making it manifest that our background is chiral. 

The one-loop effective potential can be decomposed into the contributions of fields of spin $s$, taking the compact form:
\begin{align}
   \Gamma_{\text{1-loop}} = - \frac{F^2}{8\pi^2} \sum_{\rm fields} \int_0^\infty \frac{d\tau}{\tau}   \mathcal{I}_s(F \tau)
   \label{1-loop}
\end{align}
Defining the dimensionless parameter $x \equiv F\tau$, the function $\mathcal{I}_s(x)$ is a combination of the universal orbital part (related to the Todd class) and the spin-dependent part (the Chern character), given by: 
\begin{align}
&  \mathcal{I}_s(x) = 2 (-1)^{2s}   \chi_{\rm orb}(x)   \chi_{ s}(x)  \qquad {\rm where} \cr 
 &  \chi_{\rm orb}(x)  = {\rm tr_{\rm orb}}  [ e^{-\tau (-D^2)} ],  \;\;  \chi_{ s}(x)={\rm tr}_{ s} [ e^{-\tau \Sigma F }]
  \label{fun}
\end{align}
where $(-1)^{2s}$ is due to statistics of particles. 
Here, $\chi_{\rm orb}$ is the partition function associated with the orbital fluctuation operator $-D^2= H_{12} + H_{34}$, where $H_{12} ,  H_{34}$ are 
 Landau levels Hamiltonians  in the 1-2 and 3-4 planes. The orbital spectrum is:
\begin{align}
\lambda_{n_1,n_2} = 2F(n_1+1/2) + 2F(n_2+1/2)
\end{align}
where $n_i =0,1, 2, \ldots$ are the orbital spectrum labels. The density of states is $(F/2\pi)^2$ due to the double Landau level structure, and appears in the prefactor in \eqref{1-loop}. The orbital sum gives:  
\begin{align}
    \chi_{\text{orb}}(x) =  \sum_{n_1, n_2} e^{-\tau \lambda_{n_1, n_2}}  = \frac{1}{4\sinh^2(x)}  \equiv \mathcal{I}(x)
\end{align}

For a scalar field, there is no spin interaction.  For a Dirac fermion  $S_{+} \oplus S_{-}$, left-handed components see the background, but right-handed fermions do not, generating the shifts $\{-2F, 2F, 0, 0\}$ to the orbital spectrum. 
The gauge sector is the combination of the vector bosons (transforming as $S_{+} \otimes S_{-}$) and ghosts. Note that $S_{+}$ couples to the self-dual background, while $S_{-}$ merely acts as a spectator, and we must subtract off the two ghost contributions in the $(0, 0)$ representation. As a result, the trace over the Zeeman-type Hamiltonian in the self-dual background takes the form: 
\begin{align}
      \chi_0(x) &= 1 \cr
     \chi_{1/2}(x)   &=  2 \cosh(2x) + 2 \cr
         \chi_1(x)&=       4\cosh(2x) - 2 
\end{align}

Now, we can combine the orbital and spin traces using \eqref{fun} to obtain the integration kernels. For a Majorana fermion, because of the reality condition $\psi^c = \psi$, we need to take half of the Dirac fermion contribution. Hence,   
\begin{align}
    \mathcal{I}_0(x) &= +2   \mathcal{I}(x) \cr
 \mathcal{I}_{1/2}(x) &=  -  (1+ 4   \mathcal{I}(x)) \cr
    \mathcal{I}_1(x) & = +2 (2 + 2 \mathcal{I}(x)) 
    \label{beauty}
\end{align}
This decomposition is structurally beautiful because it separates the exact zero mode contribution from the non-zero mode (higher orbital) contributions.  
The small-$x$ expansion of the orbital kernel is $\mathcal{I}(x) = \frac{1}{4x^2} - \frac{1}{12} + \mathcal{O}(x^2)$, where the $\frac{1}{4x^2}$ term yields a field-independent UV divergence that needs to be subtracted off. The constant term will be identified with the orbital contributions. For convenience, we define a subtracted kernel:
$  \widehat{\mathcal{I}}(x) =  \mathcal{I}(x) - \frac{1}{4x^2}  $.
To find the perturbative 1-loop correction to the effective action, we perform the integration 
\begin{align}
  \sum_{\rm fields} \int_{1/\Lambda_{\rm UV}^2}^{1/\mu^2} \frac{d\tau}{\tau}   \widehat{\mathcal{I}}_s (F \tau) \sim \beta_0 \ln \frac{\Lambda_{\rm UV}^2}{\mu^2}  
\end{align}
The $x\rightarrow 0$ limits of $\widehat{\mathcal{I}}_s(x)$ is in direct correspondence with the UV beta functions $(\mu \gg \sqrt {F} \gg \Lambda_{\rm YM})$, yielding:
\begin{align}
&\widehat{\mathcal{I}}_s( x\rightarrow 0) = \left\{ \begin{array}{lll} 
    4 - \frac{1}{3}= \frac{11}{3}  & \qquad s=1    \\
 - 1 + \frac{1}{3}  = -\frac{2}{3}  & \qquad  s=\frac{1}{2}  \\
     - \frac{1}{6}  &  \qquad s=0  
    \end{array} \right.
\end{align} 
The one-loop corrected effective action at scales $\mu$ where $\Lambda_{\rm UV} > \mu \geq \sqrt{F}$ takes the form:
\begin{align}
 & \Gamma_{\rm eff} (\mu; F)  = \Gamma_{0} + \Gamma_{1}  \cr
  &=  F^2 \left[ \frac{1}{g^2(\Lambda_{\rm UV})} - \frac{\beta_0}{8 \pi^2} \ln \frac{\Lambda_{\rm UV}}{\mu} \right] \equiv  \frac{1}{g^2(\mu)} F^2
  \label{UVeff}
\end{align}
which correctly reproduces the standard beta function in the deep UV:   
\begin{align}
    \beta_0&= \left(\frac{11}{3} -\frac{2}{3} n_{f}^m -\frac{1}{6} n_{sc}^{\rm r} \right)N
    \label{UV}
\end{align}
 
We can evaluate $\Gamma_{\rm eff} (\mu; F)$ at $\mu = \sqrt{F}$, as is done traditionally in the QCD literature \cite{Savvidy:1977as, Nielsen:1978rm, Nielsen:1980sx, Leutwyler:1980ma}, and write the effective action solely in terms of the background: 
\begin{align}
    \Gamma_{\rm eff} (F ) =\Gamma_{\rm eff} (\mu =\sqrt{F}), \; \; 
    g^2(F)= \frac{\beta_0}{8 \pi^2} \ln \frac{\sqrt F}{\Lambda_{\rm YM}}  
\label{BvsL}
\end{align}
Since the background is a coordinate  parameterizing  the superselection sector (a classical moduli space), the inverse running coupling naturally emerges as a metric on this field space, where $ {d^2 \Gamma(F)/d F^2}  \approx {2/g^2(\sqrt{F})}$ in the weak-coupling regime ($F \gg \Lambda_{\rm YM}^2$), similar to the metric on the moduli space in  $\mathcal{N}=2$ Seiberg-Witten theory. 
Below, we examine the structure of the theory at energy scales $\mu < \sqrt{F}$ keeping $F$ fixed.

\subsection{Abelianized theory and quantized beta function}  
In the self-dual background field, as we stated, the charged degrees of freedom sits in discrete Landau levels, while the Cartan components have continuous spectrum.  For  fermions and gauge bosons, the Landau levels have exact zero modes and  higher non-zero modes and 
for scalars only non-zero modes.   
The existence of the zero modes is in sharp contradistinction with the  adjoint Higgs mechanism where the theory also abelianizes \cite{Seiberg:1994rs, Lerche:1996xu}.

In the energy range 
$\mu \leq \sqrt{F}$, the theory abelianizes to $U(1)$ and   the higher Landau levels decouple from the dynamics. One may naively expect that the theory lands on trivial Coulomb phase. 
However, this is not so, because 
the zero modes do not decouple from the dynamics. To see the perturbative manifestation of this, let us continue running to scales much lower than $\sqrt{F}$, but still much larger than $\Lambda_{YM}$, hence, the one loop analysis is still reliable.  
For $ \mu \lesssim   \sqrt{F} $,  the integral splits to two parts:
 \begin{align}
 \qquad    
 \int_{1/\Lambda_{UV}^2}^{1/\mu^2}  &= \int_{1/\Lambda_{UV}^2}^{1/F}  \qquad +  \qquad  \int_{1/F}^{1/\mu^2}   \cr 
 & \propto  \beta_0 \log \frac{\Lambda_{UV}^2}{F}  +  \widetilde \beta_0 \log \frac{F}{\mu^2} 
 \label{split}
 \end{align}
Note that in the domain of second integration, 
the higher orbitals decouple. Hence, 
 the regularized kernel is saturated by the contribution of the zero modes. This corresponds to the paramagnetism induced by gauge bosons and fermions, with no scalar contributions.  As a result, in the  $\sqrt {F} > \mu \gg \Lambda_{YM}$ regime:
\begin{align}
& \widehat{\mathcal{I}_s}( 1 \ll x < \frac{F}{\mu^2} ) = \left\{ \begin{array}{lll} 
    4 & \qquad s=1    \\
   -1 &  \qquad s=\frac{1}{2}  \\
     0  &  \qquad s=0  
    \end{array} \right.
\end{align} 
Hence, beta function coefficient becomes:
 \begin{align}
    \widetilde \beta_0 = (4-n_{f}^m)N
    \label{running3}
 \end{align} 
 in contrast with \eqref{UV}.
The  coefficients are now quantized, there are no longer factors of 
$\frac{1}{6}$ and  $\frac{1}{3}$ that arise from the orbital sums. 

These integers are a consequence of the Atiyah-Singer index theorem in the constant self-dual background. While this connection to the Dirac operator is obvious for fermions, it also holds for the gauge sector because the linearized self-duality fluctuations, combined with gauge-fixing, precisely match a twisted Dirac operator. Although the overall dimension of the moduli space is infinite, the local index density (zero modes per unit volume) is exact: $4N$ for the gauge sector and correspondingly fixed integer densities for fermions, which together strictly determine the quantized beta function, similar to certain supersymmetric theories where one-loop beta function is  determined purely by the
instanton zero modes via the index theorem \cite{Novikov:1983uc}.

In \eqref{split},  the first term  can be combined with the  bare coupling    $g^2(\Lambda_{UV}$ )  to define  $g^2(\sqrt{F})$. 
 The effective action in this intermediate energy regime is (compare with \eqref{UVeff}):
 \begin{align}
    \Gamma(\mu) &= \left[ \frac{1}{g^2(F)} - \frac{\tilde \beta_0}{8 \pi^2} \ln\left(\frac{{\sqrt{F}}}{\mu}\right) \right] F^2 \equiv \frac{1}{g^2(\mu)} F^2 
\label{running2}
\end{align}
where $1/g^2(\sqrt{F})$ encapsulates the frozen history of the non-Abelian running. 
$\sqrt{F}$ plays the role of UV-cutoff from the point of view of low energy effective theory that we wish to construct. 
The new coefficient $\tilde{\beta}_0$ is determined by the index theorem.  

Two  interesting implications of  \eqref{running3}  are the following:  
The $n_f=4,5$ theories become IR-free Coulomb phase rather than IR-conformal as it is in the theory without background.   Thus, the IR-theory must undergo a phase transition once $F$ becomes sufficiently large.  This implies that, the $n_f=4,5$ theories,  enjoys an almost counter-intuitive 
 asymptotic and infrared freedom at two ends of the energy scale!  
Also, for theories which  do develop a strong scale, we obtain 
\begin{align}
    \widetilde \Lambda_{YM} = \left( \frac{\Lambda_{YM}}{\sqrt{F}} \right)^{\frac{\beta_0 - \widetilde \beta_0}{\widetilde \beta_0}} \Lambda_{YM} = \left( \frac{\Lambda_{YM}}{\sqrt{F}} \right)^{\frac{n_f-1}{3(4-n_f)}} \Lambda_{YM}
\end{align}
where   $\widetilde \Lambda_{YM}$ is the strong scale of the theory in  large selfdual background.

\subsection{$SU(N)$ case} 

In this section, we generalize our results from the $SU(2)$ gauge theory to the $SU(N)$ case in a self-dual background. We consider a generic (traceless) background field:
\begin{equation}
  F = \mathrm{diag}(F_1, \ldots, F_N).
  \label{generic}
\end{equation}
The off-diagonal $W_{i}^{j}$ components of the adjoint fields experience the relative field strengths $F_{ij} = F_i - F_j$. Consequently, they form $N^2-N$ Landau levels with energy gaps proportional to $2|F_{ij}|$. In contrast, the neutral components (the $U(1)^{N-1}$ Cartan photons) retain a continuous spectrum. Because the background naturally defines a rank-$(N-1)$ matrix—generalizing the rank-1 case of $SU(2)$—the effective action $\Gamma_{\rm eff}(F_i)$, acting as the counterpart to \eqref{BvsL}, is given by:
 \begin{align}
 \Gamma_{\rm eff} (F_i) 
 &=  \frac{2}{N} \sum_{ i < j} (F_{ij})^2\left( \frac{1}{g_0^2 }    -  \frac{1}{16\pi^2}   \beta_0 
 \log \frac{\Lambda_{\rm UV}^2}{|F_{ij}|}  \right).  
 \label{EffAct}
   \end{align}

Unlike the deep UV, where the renormalization group running is governed by a single universal gauge coupling $g^2(\mu)$, the $|F_{ij}|$  multi-scale  thresholds now require the introduction of a generalized $N \times N$ coupling matrix, $\tau_{ij}^\star$, of 
rank-$(N-1)$. Remarkably, the form of the effective action at 1-loop order in terms of the background fields is algebraically identical to the 1-loop Seiberg-Witten prepotential formulated in terms of adjoint Higgs vacuum expectation values on the Coulomb branch \cite{Seiberg:1994rs, Lerche:1996xu}. We can rewrite \eqref{EffAct} as:
\begin{align}
\widetilde \Gamma_{\rm eff} (F_i) &= \sum_{ i, j} {\rm Im}\left(\frac{\tau^{\star}}{2\pi}\right)_{ij}   \; F_i \;  F_j.  
\label{EffActsus}
\end{align}
   
The coupling matrix $\tau_{ij}^\star$ acts as the exact initial condition for the lower-energy theory, completely capturing the  decoupling of the non-zero modes. Because each $U(1)$ sector inherits a distinct initial condition set by its specific  threshold, this flow naturally generates a full, non-trivial generalized coupling matrix. 

Finally, if we consider the structure of the effective action at scales $\Lambda_{\rm YM} \ll \mu < \sqrt{|F_{ij}|}$ for all $(i,j)$, the running in this range is governed entirely by the topologically quantized coefficient $\widetilde \beta_0$. The effective action safely below all mass thresholds becomes: 
\begin{align}
& \Gamma_{\rm eff} (F_i; \mu)  \cr
 &  =  \frac{2}{N} \sum_{ i < j} (F_{ij})^2\left( \frac{1}{g_0^2 }    -  \frac{1}{16\pi^2}   \beta_0 
 \log \frac{\Lambda_{\rm UV}^2}{|F_{ij}|}   -  \frac{1}{16\pi^2}   \widetilde \beta_0 
 \log \frac{|F_{ij}|}{\mu^2}      \right)   \nonumber \\
 &\equiv \sum_{ i, j} {\rm Im}\left(\frac{\tau (\mu)}{2\pi}\right)_{ij}   \; F_i \;  F_j.  
\label{EffAct_IR}
\end{align}
In particular, unlike the Seiberg-Witten construction where the running stops for $\mu < |a_i - a_j|$ for all $(i, j)$, in our case, it continues via the topological beta function.

\subsection{A conjecture and many open questions}

How can an $SU(2)$ gauge theory dynamically broken to $U(1)$---ostensibly a simple theory of  photons---continue to run and exhibit asymptotic freedom? What is the exact structure of the effective field theory (EFT) at scales $\mu \lesssim \sqrt{F}$? Surely, it cannot be pure Maxwell theory. Recall that when $SU(2) \rightarrow U(1)$ via a standard adjoint Higgs field, the coupling strictly stops running below the scale of abelianization. In our case, however, we are left with a highly non-trivial theory of  pure photons!

One may wonder how can a non-trivial, asymptotically free interacting theory of photons may emerge. For example, 3-photon vertex in standard QED is zero by charge-conjugation (Furry's theorem). 
However, in our EFT, the  3-photon vertex is certainly non-trivial. 
 The W-boson Landau level zero mode wave functions are highly localized within a size, $1/\sqrt{F}$.  Imagine a $W^{+}$ zero mode  going on the loop, clock-wise and counter-clockwise. The Aharonov-Bohm phase it acquires is given in terms of projected area $A$ onto 12 and 34 planes, and field strength, and induce a phase factor for the 3-photon vertex: 
 \begin{align}
    e^{i \Phi_{\circlearrowleft}} - e^{i \Phi_{\circlearrowright}} = 2 i \sin (A F)  
 \end{align}
 If we consider  photons with external momenta $k_1, k_2, k_3$, 
the Fourier transform of the Aharonov-Bohm  phase factor enters to the 3-point amplitude in our EFT, and remarkably,  it is  the 
 Moyal star-product, 
$2i \sin\left( \frac{1}{2} k_1^\mu \Theta_{\mu\nu} k_2^\nu \right)$ where 
$\Theta_{\mu\nu} = 1/F_{\mu \nu}$, localized to an effective vertex of size $1/\sqrt{F}$. 
Similar structure also holds for 4-photon vertex.

Our conjecture is that, the EFT at our disposal must be related to non-commutative (NC) $U(1)$ theory.  However, the standard $U(1)$ NC theory is a pathological theory due to UV-IR mixing \cite{Minwalla:1999px, Ruiz:2000hu}. Deep UV manifests itself in the IR as non-locality (which is also tachyonic for $n_f=0$)   in the photon dispersion relation  $E^2(p)= p^2 - \frac{c (1-n_f)}{(\Theta p)^2 }, \; c>0$, violating a fundamental principle  of   quantum field theory \cite{Matusis:2000jf, Armoni:2001uw}.   

In this sense, the standard $U(1)$ NC theory, despite the fact that it is asymptotically free, and as such, it was initially hoped to be a UV-complete theory, is actually a sick theory due to UV-IR mixing. It violates locality.  Our construction walks around this long standing impasse.

In our construction, at energies much larger than $\sqrt{F}$, the theory has a healthy UV completion, which is non-abelian $SU(2)$ theory.   Hence, the notorious UV-IR mixing problem does not exist in our set-up. Our construction achieves for gauge theory what Grosse and Wulkenhaar achieved in scalar field theory \cite{Grosse:2004yu}.  

We should also point out  that if we start with $SU(N)$ gauge theory in a  selfdual background   \eqref{generic}, we obtain 
$U(1)^{N-1}$ generalization of the $U(1)$ non-commutative theory with 
rank $N-1$ non-commutativity matrix: 
\begin{align}
    \Theta_{\mu \nu}^{ij} \sim \frac{1}{F_{\mu \nu}^{ij}} 
\end{align}
In this way, our construction also differs from the stringy construction of Seiberg and Witten \cite{Seiberg:1999vs} where there is a unique non-commutative parameter.

This setup opens the door to several deep questions regarding the vacuum structure of Yang-Mills theory, establishing a framework to answer them. Does Yang-Mills theory in a large self-dual background exhibit confinement? Does it generate a mass gap? Naively, one might expect the answer to be an obvious ``yes,'' given that the theory is driven to strong coupling even faster. 

However, we argue that the answer is highly non-trivial because the background field induces a profound structural shift beyond simply accelerating the running of the gauge coupling $g^2(\mu)$. In our EFT, the standard instanton size moduli problem of Yang-Mills theory is dynamically stabilized by the constant self-dual background, rendering non-perturbative aspects analytically calculable. In the long-distance effective theory, we anticipate the emergence of Nekrasov-Schwarz anti-instantons of fixed size \cite{Nekrasov:1998ss, Furuuchi:2000vc, Chu:2001cx}. 


Crucially, even as the perturbative sector is driven to strong coupling, the anti-instanton sector remains strictly dilute, governed by a weight factor $\exp[- 8 \pi^2/g^2(\sqrt{F})] \ll 1$, and theoretically under control. This dynamical scale separation is strikingly reminiscent of 4d ${\mathbb Z}_N$ lattice gauge theories with a monopole suppression term, which is  shown to exhibit a strongly coupled Coulomb phase \cite{Nguyen:2024ikq}. 

Hence, we expect that the non-perturbative topological dynamics of the theory are parametrically isolated from the perturbative strong-coupling regime! Ultimately, this setup yields new, analytically controlled opportunities to understand strongly coupled gauge dynamics directly on $\mathbb{R}^4$.

\acknowledgments
We thank Jose Barbon, Philip Argyres, \"Ozg\"ur  Oktel, Mikhail Shifman, and  Arkady Vainshtein   for various conversations about this work. 
 M.\"U. is supported by U.S. Department of Energy, Office of Science, Office of Nuclear Physics, under Award Number DE-FG02-03ER41260.

\bibliography{QFT-Mithat-5}

\begin{thebibliography}{26}%
\makeatletter
\providecommand \@ifxundefined [1]{%
 \@ifx{#1\undefined}
}%
\providecommand \@ifnum [1]{%
 \ifnum #1\expandafter \@firstoftwo
 \else \expandafter \@secondoftwo
 \fi
}%
\providecommand \@ifx [1]{%
 \ifx #1\expandafter \@firstoftwo
 \else \expandafter \@secondoftwo
 \fi
}%
\providecommand \natexlab [1]{#1}%
\providecommand \enquote  [1]{``#1''}%
\providecommand \bibnamefont  [1]{#1}%
\providecommand \bibfnamefont [1]{#1}%
\providecommand \citenamefont [1]{#1}%
\providecommand \href@noop [0]{\@secondoftwo}%
\providecommand \href [0]{\begingroup \@sanitize@url \@href}%
\providecommand \@href[1]{\@@startlink{#1}\@@href}%
\providecommand \@@href[1]{\endgroup#1\@@endlink}%
\providecommand \@sanitize@url [0]{\catcode `\\12\catcode `\$12\catcode `\&12\catcode `\#12\catcode `\^12\catcode `\_12\catcode `\%12\relax}%
\providecommand \@@startlink[1]{}%
\providecommand \@@endlink[0]{}%
\providecommand \url  [0]{\begingroup\@sanitize@url \@url }%
\providecommand \@url [1]{\endgroup\@href {#1}{\urlprefix }}%
\providecommand \urlprefix  [0]{URL }%
\providecommand \Eprint [0]{\href }%
\providecommand \doibase [0]{https://doi.org/}%
\providecommand \selectlanguage [0]{\@gobble}%
\providecommand \bibinfo  [0]{\@secondoftwo}%
\providecommand \bibfield  [0]{\@secondoftwo}%
\providecommand \translation [1]{[#1]}%
\providecommand \BibitemOpen [0]{}%
\providecommand \bibitemStop [0]{}%
\providecommand \bibitemNoStop [0]{.\EOS\space}%
\providecommand \EOS [0]{\spacefactor3000\relax}%
\providecommand \BibitemShut  [1]{\csname bibitem#1\endcsname}%
\let\auto@bib@innerbib\@empty
\bibitem [{\citenamefont {'t~Hooft}(1980)}]{tHooft:1979rat}%
  \BibitemOpen
  \bibfield  {author} {\bibinfo {author} {\bibfnamefont {G.}~\bibnamefont {'t~Hooft}},\ }in\ \href {https://doi.org/10.1007/978-1-4684-7571-5_9} {\emph {\bibinfo {booktitle} {{Recent Developments in Gauge Theories. Proceedings, Nato Advanced Study Institute, Cargese, France, August 26 - September 8, 1979}}}},\ Vol.~\bibinfo {volume} {59}\ (\bibinfo {year} {1980})\ pp.\ \bibinfo {pages} {135--157}\BibitemShut {NoStop}%
\bibitem [{\citenamefont {Gaiotto}\ \emph {et~al.}(2017)\citenamefont {Gaiotto}, \citenamefont {Kapustin}, \citenamefont {Komargodski},\ and\ \citenamefont {Seiberg}}]{Gaiotto:2017yup}%
  \BibitemOpen
  \bibfield  {author} {\bibinfo {author} {\bibfnamefont {D.}~\bibnamefont {Gaiotto}}, \bibinfo {author} {\bibfnamefont {A.}~\bibnamefont {Kapustin}}, \bibinfo {author} {\bibfnamefont {Z.}~\bibnamefont {Komargodski}},\ and\ \bibinfo {author} {\bibfnamefont {N.}~\bibnamefont {Seiberg}},\ }\href {https://doi.org/10.1007/JHEP05(2017)091} {\bibfield  {journal} {\bibinfo  {journal} {JHEP}\ }\textbf {\bibinfo {volume} {05}},\ \bibinfo {pages} {091}},\ \Eprint {https://arxiv.org/abs/1703.00501} {arXiv:1703.00501 [hep-th]} \BibitemShut {NoStop}%
\bibitem [{\citenamefont {Seiberg}\ and\ \citenamefont {Witten}(1994)}]{Seiberg:1994rs}%
  \BibitemOpen
  \bibfield  {author} {\bibinfo {author} {\bibfnamefont {N.}~\bibnamefont {Seiberg}}\ and\ \bibinfo {author} {\bibfnamefont {E.}~\bibnamefont {Witten}},\ }\href {https://doi.org/10.1016/0550-3213(94)90124-4} {\bibfield  {journal} {\bibinfo  {journal} {Nucl. Phys.}\ }\textbf {\bibinfo {volume} {B426}},\ \bibinfo {pages} {19} (\bibinfo {year} {1994})},\ \bibinfo {note} {[Erratum: Nucl. Phys.B430,485(1994)]},\ \Eprint {https://arxiv.org/abs/hep-th/9407087} {arXiv:hep-th/9407087 [hep-th]} \BibitemShut {NoStop}%
\bibitem [{\citenamefont {Polyakov}(1975)}]{Polyakov:1975rs}%
  \BibitemOpen
  \bibfield  {author} {\bibinfo {author} {\bibfnamefont {A.~M.}\ \bibnamefont {Polyakov}},\ }\href {https://doi.org/10.1016/0370-2693(75)90162-8} {\bibfield  {journal} {\bibinfo  {journal} {Phys. Lett.}\ }\textbf {\bibinfo {volume} {B59}},\ \bibinfo {pages} {82} (\bibinfo {year} {1975})}\BibitemShut {NoStop}%
\bibitem [{\citenamefont {Unsal}(2009)}]{Unsal:2007jx}%
  \BibitemOpen
  \bibfield  {author} {\bibinfo {author} {\bibfnamefont {M.}~\bibnamefont {Unsal}},\ }\href {https://doi.org/10.1103/PhysRevD.80.065001} {\bibfield  {journal} {\bibinfo  {journal} {Phys. Rev.}\ }\textbf {\bibinfo {volume} {D80}},\ \bibinfo {pages} {065001} (\bibinfo {year} {2009})},\ \Eprint {https://arxiv.org/abs/0709.3269} {arXiv:0709.3269 [hep-th]} \BibitemShut {NoStop}%
\bibitem [{\citenamefont {Unsal}(2008)}]{Unsal:2007vu}%
  \BibitemOpen
  \bibfield  {author} {\bibinfo {author} {\bibfnamefont {M.}~\bibnamefont {Unsal}},\ }\href {https://doi.org/10.1103/PhysRevLett.100.032005} {\bibfield  {journal} {\bibinfo  {journal} {Phys. Rev. Lett.}\ }\textbf {\bibinfo {volume} {100}},\ \bibinfo {pages} {032005} (\bibinfo {year} {2008})},\ \Eprint {https://arxiv.org/abs/0708.1772} {arXiv:0708.1772 [hep-th]} \BibitemShut {NoStop}%
\bibitem [{\citenamefont {Tanizaki}\ and\ \citenamefont {\"Unsal}(2022)}]{Tanizaki:2022ngt}%
  \BibitemOpen
  \bibfield  {author} {\bibinfo {author} {\bibfnamefont {Y.}~\bibnamefont {Tanizaki}}\ and\ \bibinfo {author} {\bibfnamefont {M.}~\bibnamefont {\"Unsal}},\ }\href {https://doi.org/10.1093/ptep/ptac042} {\bibfield  {journal} {\bibinfo  {journal} {PTEP}\ }\textbf {\bibinfo {volume} {2022}},\ \bibinfo {pages} {04A108} (\bibinfo {year} {2022})},\ \Eprint {https://arxiv.org/abs/2201.06166} {arXiv:2201.06166 [hep-th]} \BibitemShut {NoStop}%
\bibitem [{\citenamefont {Poppitz}(2022)}]{Poppitz:2021cxe}%
  \BibitemOpen
  \bibfield  {author} {\bibinfo {author} {\bibfnamefont {E.}~\bibnamefont {Poppitz}},\ }\href {https://doi.org/10.3390/sym14010180} {\bibfield  {journal} {\bibinfo  {journal} {Symmetry}\ }\textbf {\bibinfo {volume} {14}},\ \bibinfo {pages} {180} (\bibinfo {year} {2022})},\ \Eprint {https://arxiv.org/abs/2111.10423} {arXiv:2111.10423 [hep-th]} \BibitemShut {NoStop}%
\bibitem [{\citenamefont {Leutwyler}(1981)}]{Leutwyler:1980ma}%
  \BibitemOpen
  \bibfield  {author} {\bibinfo {author} {\bibfnamefont {H.}~\bibnamefont {Leutwyler}},\ }\href {https://doi.org/10.1016/0550-3213(81)90252-2} {\bibfield  {journal} {\bibinfo  {journal} {Nucl. Phys. B}\ }\textbf {\bibinfo {volume} {179}},\ \bibinfo {pages} {129} (\bibinfo {year} {1981})}\BibitemShut {NoStop}%
\bibitem [{\citenamefont {Savvidy}(1977)}]{Savvidy:1977as}%
  \BibitemOpen
  \bibfield  {author} {\bibinfo {author} {\bibfnamefont {G.~K.}\ \bibnamefont {Savvidy}},\ }\href {https://doi.org/10.1016/0370-2693(77)90759-6} {\bibfield  {journal} {\bibinfo  {journal} {Phys. Lett. B}\ }\textbf {\bibinfo {volume} {71}},\ \bibinfo {pages} {133} (\bibinfo {year} {1977})}\BibitemShut {NoStop}%
\bibitem [{\citenamefont {Nielsen}\ and\ \citenamefont {Olesen}(1978)}]{Nielsen:1978rm}%
  \BibitemOpen
  \bibfield  {author} {\bibinfo {author} {\bibfnamefont {N.~K.}\ \bibnamefont {Nielsen}}\ and\ \bibinfo {author} {\bibfnamefont {P.}~\bibnamefont {Olesen}},\ }\href {https://doi.org/10.1016/0550-3213(78)90377-2} {\bibfield  {journal} {\bibinfo  {journal} {Nucl. Phys. B}\ }\textbf {\bibinfo {volume} {144}},\ \bibinfo {pages} {376} (\bibinfo {year} {1978})}\BibitemShut {NoStop}%
\bibitem [{\citenamefont {Nielsen}(1981)}]{Nielsen:1980sx}%
  \BibitemOpen
  \bibfield  {author} {\bibinfo {author} {\bibfnamefont {N.~K.}\ \bibnamefont {Nielsen}},\ }\href {https://doi.org/10.1119/1.12565} {\bibfield  {journal} {\bibinfo  {journal} {Am. J. Phys.}\ }\textbf {\bibinfo {volume} {49}},\ \bibinfo {pages} {1171} (\bibinfo {year} {1981})}\BibitemShut {NoStop}%
\bibitem [{\citenamefont {Gorsky}\ \emph {et~al.}(1995)\citenamefont {Gorsky}, \citenamefont {Krichever}, \citenamefont {Marshakov}, \citenamefont {Mironov},\ and\ \citenamefont {Morozov}}]{Gorsky:1995zq}%
  \BibitemOpen
  \bibfield  {author} {\bibinfo {author} {\bibfnamefont {A.}~\bibnamefont {Gorsky}}, \bibinfo {author} {\bibfnamefont {I.}~\bibnamefont {Krichever}}, \bibinfo {author} {\bibfnamefont {A.}~\bibnamefont {Marshakov}}, \bibinfo {author} {\bibfnamefont {A.}~\bibnamefont {Mironov}},\ and\ \bibinfo {author} {\bibfnamefont {A.}~\bibnamefont {Morozov}},\ }\href {https://doi.org/10.1016/0370-2693(95)00723-X} {\bibfield  {journal} {\bibinfo  {journal} {Phys. Lett. B}\ }\textbf {\bibinfo {volume} {355}},\ \bibinfo {pages} {466} (\bibinfo {year} {1995})},\ \Eprint {https://arxiv.org/abs/hep-th/9505035} {arXiv:hep-th/9505035} \BibitemShut {NoStop}%
\bibitem [{\citenamefont {Lerche}(1997)}]{Lerche:1996xu}%
  \BibitemOpen
  \bibfield  {author} {\bibinfo {author} {\bibfnamefont {W.}~\bibnamefont {Lerche}},\ }\href {https://doi.org/10.1016/S0920-5632(97)00073-X} {\bibfield  {journal} {\bibinfo  {journal} {Nucl. Phys. B Proc. Suppl.}\ }\textbf {\bibinfo {volume} {55}},\ \bibinfo {pages} {83} (\bibinfo {year} {1997})},\ \Eprint {https://arxiv.org/abs/hep-th/9611190} {arXiv:hep-th/9611190} \BibitemShut {NoStop}%
\bibitem [{\citenamefont {Minwalla}\ \emph {et~al.}(2000)\citenamefont {Minwalla}, \citenamefont {Van~Raamsdonk},\ and\ \citenamefont {Seiberg}}]{Minwalla:1999px}%
  \BibitemOpen
  \bibfield  {author} {\bibinfo {author} {\bibfnamefont {S.}~\bibnamefont {Minwalla}}, \bibinfo {author} {\bibfnamefont {M.}~\bibnamefont {Van~Raamsdonk}},\ and\ \bibinfo {author} {\bibfnamefont {N.}~\bibnamefont {Seiberg}},\ }\href {https://doi.org/10.1088/1126-6708/2000/02/020} {\bibfield  {journal} {\bibinfo  {journal} {JHEP}\ }\textbf {\bibinfo {volume} {02}},\ \bibinfo {pages} {020}},\ \Eprint {https://arxiv.org/abs/hep-th/9912072} {arXiv:hep-th/9912072} \BibitemShut {NoStop}%
\bibitem [{\citenamefont {Ruiz}(2001)}]{Ruiz:2000hu}%
  \BibitemOpen
  \bibfield  {author} {\bibinfo {author} {\bibfnamefont {F.~R.}\ \bibnamefont {Ruiz}},\ }\href {https://doi.org/10.1016/S0370-2693(01)00145-9} {\bibfield  {journal} {\bibinfo  {journal} {Phys. Lett. B}\ }\textbf {\bibinfo {volume} {502}},\ \bibinfo {pages} {274} (\bibinfo {year} {2001})},\ \Eprint {https://arxiv.org/abs/hep-th/0012171} {arXiv:hep-th/0012171} \BibitemShut {NoStop}%
\bibitem [{\citenamefont {Grosse}\ and\ \citenamefont {Wulkenhaar}(2005)}]{Grosse:2004yu}%
  \BibitemOpen
  \bibfield  {author} {\bibinfo {author} {\bibfnamefont {H.}~\bibnamefont {Grosse}}\ and\ \bibinfo {author} {\bibfnamefont {R.}~\bibnamefont {Wulkenhaar}},\ }\href {https://doi.org/10.1007/s00220-004-1285-2} {\bibfield  {journal} {\bibinfo  {journal} {Commun. Math. Phys.}\ }\textbf {\bibinfo {volume} {256}},\ \bibinfo {pages} {305} (\bibinfo {year} {2005})},\ \Eprint {https://arxiv.org/abs/hep-th/0401128} {arXiv:hep-th/0401128} \BibitemShut {NoStop}%
\bibitem [{\citenamefont {Langmann}\ and\ \citenamefont {Szabo}(2002)}]{Langmann:2002cc}%
  \BibitemOpen
  \bibfield  {author} {\bibinfo {author} {\bibfnamefont {E.}~\bibnamefont {Langmann}}\ and\ \bibinfo {author} {\bibfnamefont {R.~J.}\ \bibnamefont {Szabo}},\ }\href {https://doi.org/10.1016/S0370-2693(02)01650-7} {\bibfield  {journal} {\bibinfo  {journal} {Phys. Lett. B}\ }\textbf {\bibinfo {volume} {533}},\ \bibinfo {pages} {168} (\bibinfo {year} {2002})},\ \Eprint {https://arxiv.org/abs/hep-th/0202039} {arXiv:hep-th/0202039} \BibitemShut {NoStop}%
\bibitem [{\citenamefont {Novikov}\ \emph {et~al.}(1983)\citenamefont {Novikov}, \citenamefont {Shifman}, \citenamefont {Vainshtein},\ and\ \citenamefont {Zakharov}}]{Novikov:1983uc}%
  \BibitemOpen
  \bibfield  {author} {\bibinfo {author} {\bibfnamefont {V.~A.}\ \bibnamefont {Novikov}}, \bibinfo {author} {\bibfnamefont {M.~A.}\ \bibnamefont {Shifman}}, \bibinfo {author} {\bibfnamefont {A.~I.}\ \bibnamefont {Vainshtein}},\ and\ \bibinfo {author} {\bibfnamefont {V.~I.}\ \bibnamefont {Zakharov}},\ }\href {https://doi.org/10.1016/0550-3213(83)90338-3} {\bibfield  {journal} {\bibinfo  {journal} {Nucl. Phys. B}\ }\textbf {\bibinfo {volume} {229}},\ \bibinfo {pages} {381} (\bibinfo {year} {1983})}\BibitemShut {NoStop}%
\bibitem [{\citenamefont {Matusis}\ \emph {et~al.}(2000)\citenamefont {Matusis}, \citenamefont {Susskind},\ and\ \citenamefont {Toumbas}}]{Matusis:2000jf}%
  \BibitemOpen
  \bibfield  {author} {\bibinfo {author} {\bibfnamefont {A.}~\bibnamefont {Matusis}}, \bibinfo {author} {\bibfnamefont {L.}~\bibnamefont {Susskind}},\ and\ \bibinfo {author} {\bibfnamefont {N.}~\bibnamefont {Toumbas}},\ }\href {https://doi.org/10.1088/1126-6708/2000/12/002} {\bibfield  {journal} {\bibinfo  {journal} {JHEP}\ }\textbf {\bibinfo {volume} {12}},\ \bibinfo {pages} {002}},\ \Eprint {https://arxiv.org/abs/hep-th/0002075} {arXiv:hep-th/0002075} \BibitemShut {NoStop}%
\bibitem [{\citenamefont {Armoni}\ and\ \citenamefont {Lopez}(2002)}]{Armoni:2001uw}%
  \BibitemOpen
  \bibfield  {author} {\bibinfo {author} {\bibfnamefont {A.}~\bibnamefont {Armoni}}\ and\ \bibinfo {author} {\bibfnamefont {E.}~\bibnamefont {Lopez}},\ }\href {https://doi.org/10.1016/S0550-3213(02)00290-0} {\bibfield  {journal} {\bibinfo  {journal} {Nucl. Phys. B}\ }\textbf {\bibinfo {volume} {632}},\ \bibinfo {pages} {240} (\bibinfo {year} {2002})},\ \Eprint {https://arxiv.org/abs/hep-th/0110113} {arXiv:hep-th/0110113} \BibitemShut {NoStop}%
\bibitem [{\citenamefont {Seiberg}\ and\ \citenamefont {Witten}(1999)}]{Seiberg:1999vs}%
  \BibitemOpen
  \bibfield  {author} {\bibinfo {author} {\bibfnamefont {N.}~\bibnamefont {Seiberg}}\ and\ \bibinfo {author} {\bibfnamefont {E.}~\bibnamefont {Witten}},\ }\href {https://doi.org/10.1088/1126-6708/1999/09/032} {\bibfield  {journal} {\bibinfo  {journal} {JHEP}\ }\textbf {\bibinfo {volume} {09}},\ \bibinfo {pages} {032}},\ \Eprint {https://arxiv.org/abs/hep-th/9908142} {arXiv:hep-th/9908142} \BibitemShut {NoStop}%
\bibitem [{\citenamefont {Nekrasov}\ and\ \citenamefont {Schwarz}(1998)}]{Nekrasov:1998ss}%
  \BibitemOpen
  \bibfield  {author} {\bibinfo {author} {\bibfnamefont {N.}~\bibnamefont {Nekrasov}}\ and\ \bibinfo {author} {\bibfnamefont {A.}~\bibnamefont {Schwarz}},\ }\href {https://doi.org/10.1007/s002200050496} {\bibfield  {journal} {\bibinfo  {journal} {Commun. Math. Phys.}\ }\textbf {\bibinfo {volume} {198}},\ \bibinfo {pages} {689} (\bibinfo {year} {1998})},\ \Eprint {https://arxiv.org/abs/hep-th/9802068} {arXiv:hep-th/9802068} \BibitemShut {NoStop}%
\bibitem [{\citenamefont {Furuuchi}(2001)}]{Furuuchi:2000vc}%
  \BibitemOpen
  \bibfield  {author} {\bibinfo {author} {\bibfnamefont {K.}~\bibnamefont {Furuuchi}},\ }\href {https://doi.org/10.1143/PTPS.144.79} {\bibfield  {journal} {\bibinfo  {journal} {Prog. Theor. Phys. Suppl.}\ }\textbf {\bibinfo {volume} {144}},\ \bibinfo {pages} {79} (\bibinfo {year} {2001})},\ \Eprint {https://arxiv.org/abs/hep-th/0010006} {arXiv:hep-th/0010006} \BibitemShut {NoStop}%
\bibitem [{\citenamefont {Chu}\ \emph {et~al.}(2002)\citenamefont {Chu}, \citenamefont {Khoze},\ and\ \citenamefont {Travaglini}}]{Chu:2001cx}%
  \BibitemOpen
  \bibfield  {author} {\bibinfo {author} {\bibfnamefont {C.-S.}\ \bibnamefont {Chu}}, \bibinfo {author} {\bibfnamefont {V.~V.}\ \bibnamefont {Khoze}},\ and\ \bibinfo {author} {\bibfnamefont {G.}~\bibnamefont {Travaglini}},\ }\href {https://doi.org/10.1016/S0550-3213(01)00576-4} {\bibfield  {journal} {\bibinfo  {journal} {Nucl. Phys. B}\ }\textbf {\bibinfo {volume} {621}},\ \bibinfo {pages} {101} (\bibinfo {year} {2002})},\ \Eprint {https://arxiv.org/abs/hep-th/0108007} {arXiv:hep-th/0108007} \BibitemShut {NoStop}%
\bibitem [{\citenamefont {Nguyen}\ \emph {et~al.}(2025)\citenamefont {Nguyen}, \citenamefont {Sulejmanpasic},\ and\ \citenamefont {{\"U}nsal}}]{Nguyen:2024ikq}%
  \BibitemOpen
  \bibfield  {author} {\bibinfo {author} {\bibfnamefont {M.}~\bibnamefont {Nguyen}}, \bibinfo {author} {\bibfnamefont {T.}~\bibnamefont {Sulejmanpasic}},\ and\ \bibinfo {author} {\bibfnamefont {M.}~\bibnamefont {{\"U}nsal}},\ }\href {https://doi.org/10.1103/PhysRevLett.134.141902} {\bibfield  {journal} {\bibinfo  {journal} {Phys. Rev. Lett.}\ }\textbf {\bibinfo {volume} {134}},\ \bibinfo {pages} {141902} (\bibinfo {year} {2025})},\ \Eprint {https://arxiv.org/abs/2401.04800} {arXiv:2401.04800 [hep-th]} \BibitemShut {NoStop}%
\end{thebibliography}%

\end{document}